\documentclass[
reprint,
longbibliography,
 amsmath,amssymb,
 aps,
 prd,
]{revtex4-2}
\usepackage{graphicx}% Include figure files
\usepackage{verbatim}% comment
\usepackage[hidelinks]{hyperref}% hyperref
\usepackage{xcolor}% comment

\definecolor{darkblue}{rgb}{0.0, 0.1, 0.6}

\hypersetup{
  colorlinks   = true, %Colours links instead of ugly boxes
  urlcolor     = darkblue, %Colour for external hyperlinks
  linkcolor    = darkblue, %Colour of internal links
  citecolor    = darkblue %Colour of citations
}

\begin{document}
\title{Universality of curvature invariants in critical vacuum gravitational  collapse}

\author{Tom\'a\v{s} Ledvinka}
 \email{tomas.ledvinka@mff.cuni.cz}
 
\author{Anton Khirnov}%
 \email{anton@khirnov.net}

\affiliation{%
Institute of Theoretical Physics, Faculty of Mathematics and Physics, Charles University, CZ-180 00 Prague, Czech Republic
}

\begin{abstract}
We report on a numerical study of gravitational waves undergoing  gravitational collapse due to their self-interaction. We consider several families of  asymptotically flat initial data which, similar to the well-known Choptuik's discovery, can be fine-tuned between dispersal into empty space and collapse into a black hole. We find that near-critical spacetimes exhibit behavior similar to scalar-field collapse: For different families of initial data, we observe universal ``echoes'' in the form of irregularly repeating, approximate, scaled copies of the same piece of spacetime.
\end{abstract}

\maketitle

\section{\label{sec:Introduction}Introduction}
Choptuik's surprising discovery of critical behavior in gravitational collapse \cite{Choptuik_1993} showed that numerical simulations of the Einstein equations may reveal unforeseen features of the theory.
It turned out that evolution of a spherically symmetric massless scalar field minimally coupled to general relativity with initial data (ID) near the threshold between field dispersal and collapse into a black hole takes the form of ``echoes'' --- repeated concentrations of the field appearing on progressively  smaller scales. While the first few echoes retain some imprint of the ID family, later on the metric and the field look like scaled-down copies of earlier moments (discrete self-similarity) and approach the same  profile for any ID fine-tuned towards the black-hole threshold (universality).
Extensive follow-up research (see Ref. \cite{GundlachLRR} for a detailed review) then showed that similar behavior occurs for many other fields.

If the initial data form a one-parameter family, with $A$ being the parameter, then quantities such as the mass of the black hole created during the collapse $m_{\rm BH}$, the number of echoes observed, or the maximal field strength  depend on $A$ in a characteristic way as it approaches the critical value $A_*$; e.g., $m_{\rm BH}\sim (A-A_*)^\gamma$ with $\gamma\doteq 0.373$ for spherically symmetric minimally coupled massless scalar field. Later on, more detailed features  such as a periodic modulation \cite{GundlachLRR} and a lower bound of the scaling law for $m_{\rm BH}$ \cite{Puerrer_2005} were described.

Gravitational waves (GWs), an appealing alternative to a massless scalar field, have also been studied in the context of critical collapse
\cite{Abrahams_1993,Abrahams_1994,alcubierre_2000,garfinkle_2001,rinne_2008,Sorkin_2011,Hilditch_2013,Hilditch_2016,Hilditch_2017}
(see Ref. \cite{Hilditch_2017} for a detailed description of these  attempts).

In general relativity (in 3+1 dimensions; cf. \cite{Bizon2005}), there are no spherically symmetric gravitational waves, which makes numerical simulations much more computationally demanding.
Even with modern powerful computers and advanced numerical techniques, no universal and across scales repeating profile of the gravitational field analogous to that in the seminal paper \cite{Choptuik_1993} has been reported.
Moreover, Hilditch et al. \cite{Hilditch_2017} brought a strong argument against a simple  analogy with spherically symmetric collapse --- for the initial data closest to the critical amplitude, a pair of apparent horizons appeared.
Recently, critical collapse away from spherical symmetry has also been probed using models of combined gravitational and electromagnetic fields \cite{Hilditch_2019} and a semilinear scalar wave \cite{Hilditch_2020}.

\vspace{1em}
\section{Methods}
Our numerical simulations use an unconstrained evolution scheme, the so-called  Baumgarte-Shapiro-Shibata-Nakamura version of the Einstein equations \cite{Shibata_1995,Baumgarte_1998}. We use the Einstein Toolkit framework \cite{CactusCode2012} modified to analytically assume axial symmetry, and extended by a code solving elliptic equations for initial data and slicing.

{\it Coordinate choice:~} The standard 1+log slicing condition has been shown to break down in the numerical evolution of collapsing gravitational waves \cite{Hilditch_2013}.
Instead, we use the quasimaximal slicing (QMS) that we introduced recently \cite{AKTL2018} to handle such highly dynamic spacetime geometries.
We changed its implementation to use a multi-grid method together with a scheme based on Ref. \cite{PretoriusChoptuik2006} in order to be compatible with the Berger-Oliger mesh refinement, which we use to evolve hyperbolic equations.

{\it Initial data:~}
In the 3+1 approach to general relativity, initial data are specified as 3-tensor fields $\gamma_{ij}$ and $K_{ij}$ --- the intrinsic metric and the extrinsic curvature --- on the initial Cauchy hypersurface. These fields must satisfy a set of coupled nonlinear elliptic equations (Hamilton and momentum constraints) implied by the Einstein equations.
We study two axi- and plane-symmetric ID families with trace $K=\gamma^{ij}K_{ij}=0$ (i.e., compatible with  the maximal slicing).
The first family is the Brill data, first studied numerically in Ref. \cite{Eppley_1977}. Their main feature is time-symmetry due to $K_{ij}(t=0)=0$.
Gravitational waves are then encoded as a deformation of the initial slice intrinsic metric $\gamma_{ij}(t=0)$ written in standard spherical coordinates,
\begin{equation}
\gamma_{ij}dx^i\,dx^j = \psi^4\left[e^{2 q} (dr^2+r^2 d\theta^2)+r^2\sin^2\theta d\phi^2\right].
\label{BrillInit}
\end{equation}
We choose the ``seed function'' thoroughly studied in Ref. \cite{Hilditch_2017}
\begin{equation}
q(x^i) = A\; \sigma^{-2} r^2 e^{-r^2/\sigma^2} \sin^2 \theta,
\label{BrillSeedQ}
\end{equation}
where
we introduce a scale parameter $\sigma$, leaving $A$ dimensionless. The conformal factor $\psi$ must be found by solving the Hamilton constraint.

The second family of initial data is inspired by Ref. \cite{Abrahams_1994} where the initial 3-metric is taken to be conformally flat, and one component of $K_{ij}$ --- $K^r_{~\theta}$ --- is chosen to be the deformation seed. This leads to three coupled constraint equations to solve for $\psi$, $K^r_{~r}$ and $K^\phi_{~\phi}$. However, despite following Ref. \cite{Abrahams_1994} to the best of our ability, we were not able to reproduce their data exactly (as is clear, e.g., from very different critical amplitudes),
and so we use it merely as inspiration, choosing a visually similar profile
\begin{equation}
K^r_\theta(x^i,t=0) = A \sigma^{-4} r^2 (\sigma-r) e^{-r^2/\sigma^2} \sin 2\theta.
\label{taData}
\end{equation}
Solutions of the constraints for $A > 0$ then turn out to be non-unique in a way very similar to Ref. \cite{PfeifferYork2005}; there exists a value $A_\text{max} \approx 1.36247$ such that there are two solutions for $0 < A < A_\text{max}$ and none for $A > A_\text{max}$. On the ``lower'' branch, the data behave as expected; they approach flat space as $A\rightarrow 0$ and their Arnowitt-Deser-Misner (ADM) mass $M_\text{ADM}$ grows with increasing $A$. By contrast, on the ``upper'' branch the mass grows with \emph{decreasing} $A$, apparently diverging as $A \rightarrow 0$. As $A \rightarrow A_\text{max}$, both branches approach the same  solution, so we can consider them together as a single ID family with continuously growing ADM mass. We mark the upper-branch solutions with a bar; e.g., $A=\overline{1.0}$ is an upper-branch solution with $M_\text{ADM}\doteq1.06\sigma$, while $A=1.0$ is a lower-branch solution with $M_\text{ADM}\doteq0.104\sigma$.

One can also consider negative values of $A$ --- for ID \eqref{BrillInit} this leads to a different initial data family \cite{Hilditch_2013}. However, for ID \eqref{taData} replacing $A\rightarrow-A$ merely flips the sign of $K_{ij}$; i.e., we get the same initial slice evolved backwards in time. Since the data are time-asymmetric (TA), critical collapse can be studied for $A<0$ as for a new ID family.

{\it Coordinate-independent analysis:~} 
Even though the Kretschmann scalar $I_K \equiv R_{\alpha\beta\gamma\delta}R^{\alpha\beta\gamma\delta}  (\alpha,\beta=0..3)$ built from the Riemann tensor $R_{\alpha\beta\gamma\delta}$ is not a direct measure of spacetime curvature due to Lorentzian signature of spacetime metric, it is still an obvious coordinate-independent scalar quantity providing an invariant indicator of the gravitational field strength. In an axisymmetric vacuum spacetime, there are additional coordinate-independent scalars available. The circumferential radius $\rho$ and the norm of its gradient $\rho_{,\alpha}\rho^{,\alpha}$ are the simplest ones, but their values are trivial at the axis of symmetry $\rho=0$. We thus propose to use their combination $\zeta \equiv (1-\rho_{,\alpha}\rho^{,\alpha})/\rho^2$ (completed by an appropriate limit at the axis; see also \cite{AKTL2018})
as a coordinate-independent indicator of the spacetime geometry.
It can be shown that $\Psi_{2\;|\rho=0}=\tfrac{1}{2}\zeta_{|\rho=0}$ 
is the only non-vanishing projection of the Weyl tensor (as defined, e.g., in \cite{Stephani_Exact_solutions}) onto an axis-aligned null tetrad,
so at the axis, we also have $I_{K\;|\rho=0} = 12\zeta^2_{|\rho=0}$.

\vspace{1em}
\section{Results}

{\it Critical amplitudes: }
We observed behavior compatible with the existence of critical amplitudes separating dispersal and black-hole formation.
The limiting factor in near-critical simulations is the sufficient resolution of the QMS solver, without which the coordinate singularities known from Ref. \cite{Hilditch_2013} appear. Importantly, it turned out that some ID  families are less prone to those pathologies than others and so require less computational effort.
Among several attempts, the initial data \eqref{taData} appeared to be least demanding. We concentrated the available resources ($\sim\! 10^4$ CPU-hours per run) here and obtained five echoes and $A_*^{\rm TA+}\doteq\overline{1.3008079^{\pm 4}}$.
For negative values of the parameter, we get  $A_*^{\rm TA-}\doteq\overline{-1.22434^{\pm 5}}$.
The Brill initial data \eqref{BrillInit} defied our bisection attempts most, and we got an interval $A_*^{\rm Brill+}\doteq{4.697^{\pm 1}}$ compatible
with the much better result $A_*^{\rm Brill+}\doteq{4.6966953 ^{\pm 78}}$ in Ref. \cite{Hilditch_2017}. With negative $A$ and less effort, we found $A_*^{\rm Brill-}\doteq{-3.509106^{\pm 5}}$.

\begin{figure}
    \centering
    \includegraphics[width=0.99\linewidth]{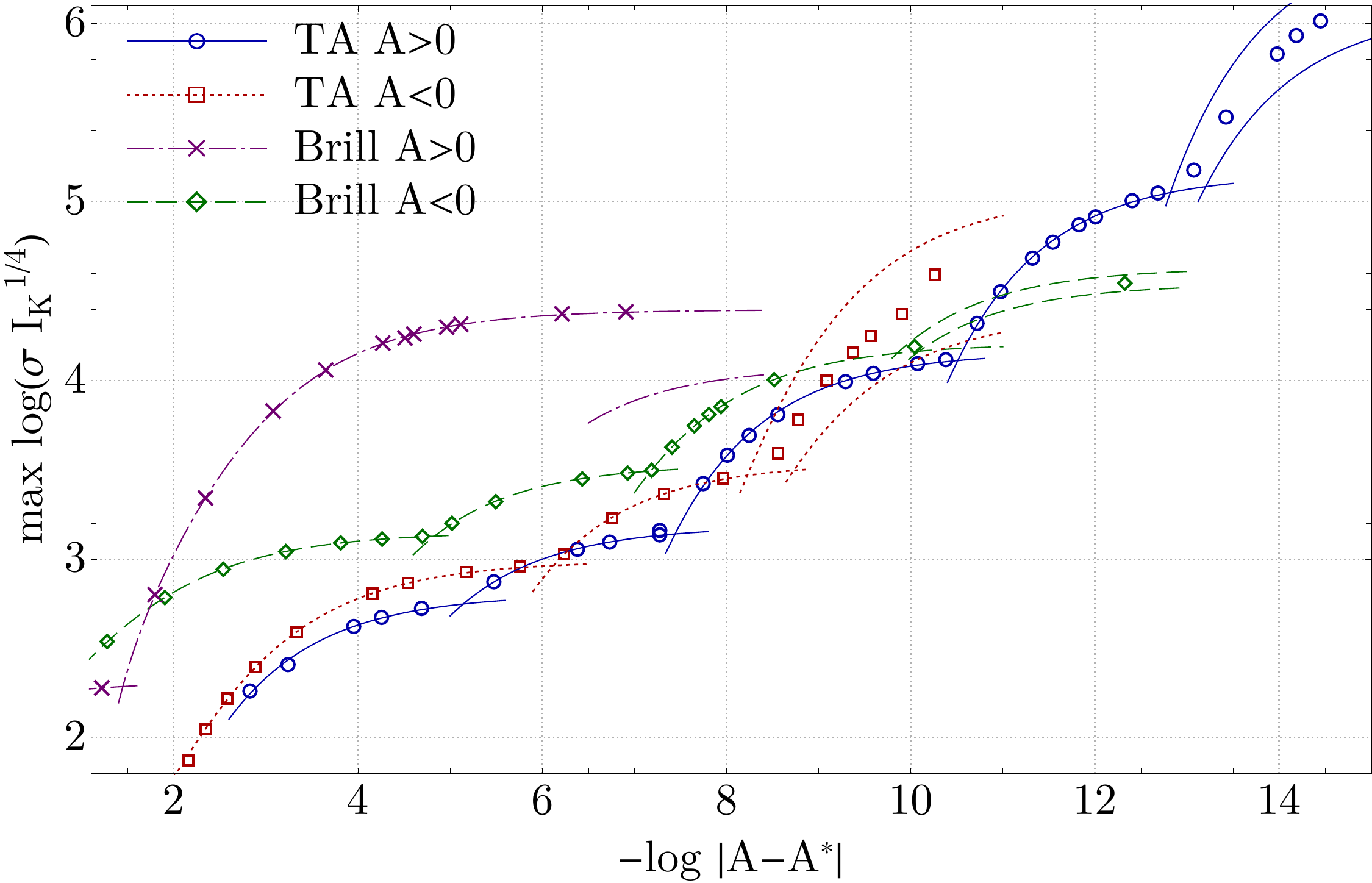}
        \caption{Global maxima of the Kretschmann invariant in subcritical spacetimes
with four families of initial data depending on a parameter $A$.
As $A$ approaches the critical value $A_*$, the maxima get ever larger, as newly appearing local extrema overtake earlier ones.
%The plotted curves are thus composed of segments --- each corresponding to a specific local maximum being the strongest one --- fitted by $\log I_K^{max} = pA +q$.
To illustrate the smooth dependence of these local extrema (echoes) on the parameter $A$, we fit the simulation results shown as points with a polynomial --- typically a simple linear dependence $\log I_K^\text{max} = p A +q$. The plotted curves are thus composed of segments, each corresponding to a specific local maximum being the strongest one.
An effect of the uncertainty of $A_*$ within the final bisection interval is indicated in the rightmost segments; $A_*^\text{Brill+}$ is taken from Ref. \cite{Hilditch_2017}.
    \label{fig:FigIKofA}
    }
\end{figure}

{\it Scaling:}
To relate our work to existing results, we start by discussing how the extrema of the Kretschmann scalar $I_K^{\rm max}$ depend
 on the amplitude parameter $A$. 
Plots showing $I_K^{\rm max}(A)$, which reduce the entire evolution of the initial data to a single number, are inspired by the typical behavior of a scale-invariant spherically-symmetric critical collapse. It admits critical solutions exhibiting a discrete self-symmetry (DSS) \cite{GundlachLRR}, i.e., containing a geometric sequence with a quotient $e^{-\Delta}$ of scaled-down copies of the same field configuration. The evolution of near-critical initial data  in the central region (in the past null cone of the accumulation point) first approaches this solution, then exhibits several almost-DSS cycles, and finally, either the scalar field undergoes dispersion or forms a black hole.
Then, for subcritical spacetimes, the quantity
$(I_K^{\rm max})^{1/4}$ with the dimension of inverse length indicates the smallest scale up to which the evolution in the central region stays close to the critical solution. The approximate relation between this scale and the parameter A again has the form $(I_K^{\rm max})^{-1/4} \sim |A-A_*|^\gamma$.

For subcritical GW collapse, we observed that near  $A_*$ the Kretschmann invariant $I_K$ attains its most pronounced extrema at the axis of symmetry coinciding with the minima of the invariant $\zeta$. As $A\rightarrow A_*$, echoes with ever higher amplitudes appear, and the strongest echo for a given $A$ determines  the value $I_K^{\rm max}$. These are shown in Fig. \ref{fig:FigIKofA}, where individual simulations are shown as data points for four families of initial data.
(The overlapping markers at $|A-A_*|\approx e^{-7.3}$
show the simulation for which $I_K^{\rm max}$ 
appears off the $z$-axis.) Each ID family can be approximated by a power law, with the critical exponent estimates $\gamma_{\text{TA}^+}=0.35^{\pm3}$, $\gamma_{\text{TA}^-}= 0.37^{\pm 8}$, and $\gamma_{\rm Brill-}=0.19^{\pm3}$. According to Ref. \cite{Hilditch_2017}, $\gamma_{\rm Brill+}\doteq 0.37$.
In our fits, we excluded data points corresponding to the first echo so that a direct influence of the initial data form is suppressed. A similar approach applied to  the results of Ref. \cite{Hilditch_2017} seems to yield $\gamma_{\rm Brill+}> 0.5$.

These differences in the exponent $\gamma$
appear to be significant, but we cannot decide if the slopes in Fig. \ref{fig:FigIKofA} really settle towards a specific value for a given family or whether they continue fluctuating significantly with further echoes appearing without apparent order.
One could claim that our simulations are merely not close enough to $A_*$, but we will show that individual echoes  have a universal form, so this argument does not seem convincing.

\begin{figure}
    \centering
    \includegraphics[width=\linewidth]{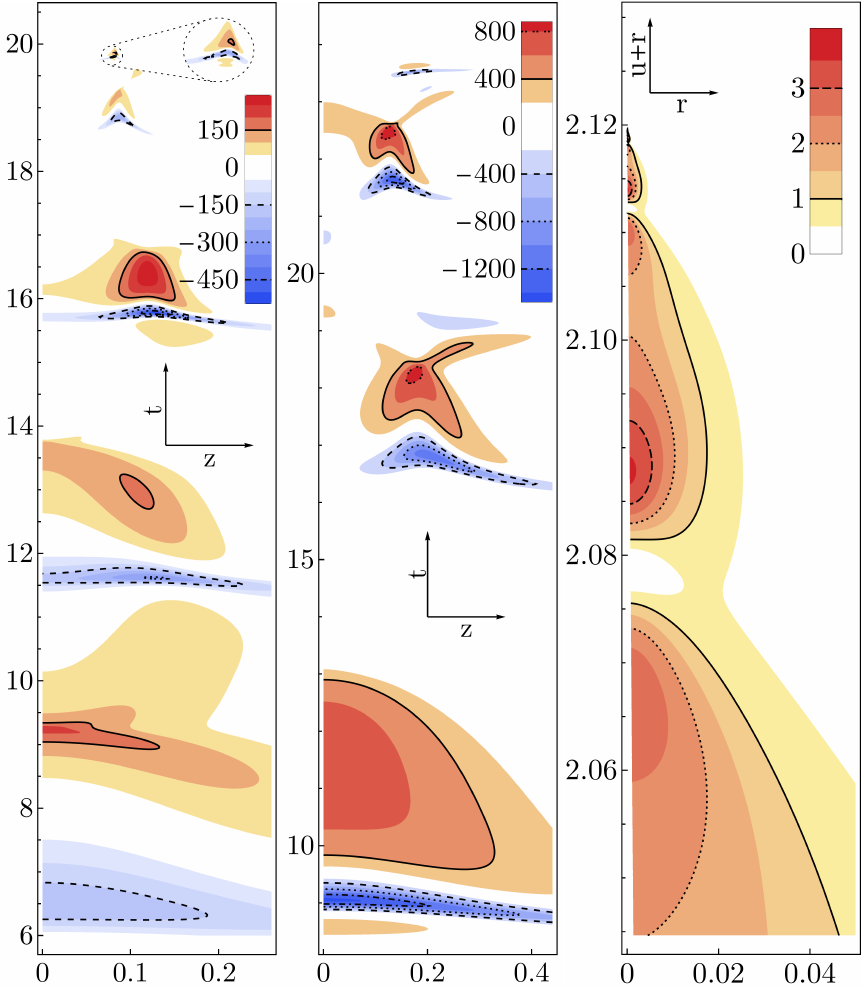}
    \caption{
Comparison of the echoes in the curvature invariant $\zeta$ between different collapse scenarios. Since its extrema span several orders of magnitude, shown are the contours of a dimensionless (but coordinate-dependent; see text) quantity $(\tau_* - \tau)^2\zeta$ in the $t-z$ plane. In a DSS setup, this quantity would repeatedly acquire the same extremal values  on ever smaller scales as $\tau \rightarrow \tau_*$ (near-critical massless scalar-field collapse \cite{Puerrer_2005} in the right panel). Left: $A_{\text{TA}^+}=\overline{1.30080828},z_0=0.08415\sigma,\tau_*=3.88\sigma$; center: $A_{\rm Brill-}={-3.5090625},z_0=0.161\sigma,\tau_*=5.9\sigma$   }
    \label{fig:tzplot}
\end{figure}

We categorize the ID as supercritical if we find an apparent horizon (AH).
Because some  AHs grow rapidly at first,
and it is impractical to store all the data for post-processing, we determine the initial AH mass $M_\text{AH}$  with a considerable ``sampling'' error. We estimate it to be $\lesssim 20\%$ for $-9.5<\epsilon_A<-2$, where  $\epsilon_A \equiv \log|A-A_*|$. In this interval  $\mu\equiv\log(M_\text{AH}/\sigma)$ satisfies
$\mu_{\text{TA}^+} = -0.44+0.17 \epsilon_A \pm 0.15$, 
$\mu_{\text{TA}^-} = 0.07 + 0.21 \epsilon_A \pm 0.13$,
and $\mu_{\text{Brill}^-}=-0.49+0.17\epsilon_A\pm 0.10$.
For ID \eqref{BrillInit} with $A>0$ we get $\gamma\approx 0.16$
for  $-6<\epsilon_A<-1$. 
Below these intervals, we observe bifurcated horizons:
For ID \eqref{BrillInit}  with $A_{{\rm Brill}^+}=4.698$, in agreement with Ref. \cite{Hilditch_2017}, we find a pair of AHs, each with $M_\text{AH} \doteq 0.10\sigma$. In addition, for ID \eqref{taData} with $A_{\text{TA}^+} = \overline{1.3008012}$ we get a pair with $M_\text{AH} \doteq 0.037\sigma$. By contrast, the $I_K^\text{max}$ slopes $\gamma_{\text{TA}^\pm}$ and $\gamma_{\rm Brill-}$ given above include only the bifurcated curvature extrema (as seen in Fig. \ref{fig:tzplot}). Thus, we assume they describe the behavior close to $A_*$ more faithfully. Further complications associated with the use of apparent horizons for critical behavior investigation are discussed in Ref \cite{Hilditch_2017}.

{\it Self-similarity and universality:}
Typically, the scale-invariant collapse in spherical symmetry   permits a DSS critical solution. An approximation of this critical solution then appears inside a so-called self-similarity horizon for an arbitrary spherically symmetric one-parameter ID family if the parameter is fine-tuned between dispersal and collapse. In this Letter, we argue that for GW, we find analogous yet more complicated behavior:
The spacetime regions near the extrema of $I_K^{\rm max}$ appear repeatedly as scaled approximate copies (a limited and irregular analog of self-similarity) of the same piece of a spacetime and independently of the ID family (universality).

In a spherically symmetric DSS spacetime and assuming adapted coordinates, all dimensionless quantities are periodic functions of the logarithmic time $\log |\tau_*-\tau|$ \cite{GundlachLRR}, with $\tau$ being the central proper time, which takes the value of $\tau_*$ at the accumulation event. Because the curvature extrema in critical GW collapse appear at $z\ne0$, to define $\tau$ we choose the worldline of constant $z=z_0$ on which the global maximum of the Kretschmann invariant appears. Because of our choice of shift $\beta^i=0$, this worldline is timelike. We then distribute $\tau$ over our (approximately maximal) slices $t=\rm const$. and construct a dimensionless quantity $(\tau_* - \tau)^2 \zeta$. Despite its coordinate dependence, it is remarkably efficient in ``equalizing'' the echoes to a common scale. The spacetime diagrams in Fig. 2 demonstrate this by showing the similarity of the echoes across very different families \eqref{BrillInit} and \eqref{taData}.

The right panel shows the same quantity for a scalar field collapse computed according to Ref. \cite{Puerrer_2005}, where we take $\tau\equiv u+r$ outside of the center.
It illustrates significant differences  between scalar-field and GW critical collapse, with the latter having spacetime curvature concentrated into irregularly appearing spikes with $\approx200\times$ larger values of the same \emph{dimensionless} quantity.

To assess the tendency towards DSS behavior, we consider consecutive local spacetime minima $\zeta_{n-1}, \zeta_n$; separated by the geodesic proper time interval $\tau_{n}-\tau_{n-1}$. We define $\Delta_n^{(\zeta)}\equiv\log[(\zeta_n/\zeta_{n-1})^{1/2}]$ for the curvature-scale ratio and $\Delta_n^{(\tau)} \equiv\log[(\tau_{n}-\tau_{n-1})/(\tau_{n+1}-\tau_n)]$ for the time-scale ratio.
In a DSS spacetime $\Delta_n^{(\zeta)}\!\!=\!\!\Delta_n^{(\tau)}\!\!=\!\!\Delta$.
From $\zeta_n$ in sub- and super-critical simulations closest to $A_*$, we obtain
$\Delta_{\text{TA}^+}^{(\zeta)}=\{0.68, 0.98, 1.02^{\pm2},1.2^{\pm3}\}$,
$\Delta_{\text{Brill}^-}^{(\zeta)}=\{0.38,0.68^{\pm1},0.5^{\pm1}\}$, and $\Delta_{\;\text{TA}^-}^{(\zeta)}=\{0.55,0.96^{\pm4}\}$
for five, four, and three echoes seen for respective ID in Fig. \ref{fig:FigIKofA}.
For the time-scales, we get
$\Delta_{\text{TA}^+}^{(\tau)}=\{-0.1,0.4,2.1^{\pm1}\}$,
$\Delta_{\text{Brill}^-}^{(\tau)}=\{0.3,1.1^{\pm1}\}$
and $\Delta_{\text{TA}^-}^{(\tau)}=\{0.1\}$.
%The spacetime function $\tau$ extended from the worldline $z=z_0$ used in Fig. \ref{fig:tzplot} yields $\Delta_{\text{TA}^+}^{(\tau)}=\{0.1,0.3,1.7\}$.

Although these numbers seem incompatible with DSS behavior, it is remarkable that while $\zeta$ spans the ratio $>400$, using the quantity $(\tau_*-\tau)^2\zeta$ devised on self-similarity arguments, this ratio reduces to a factor $\sim 3$.

To demonstrate the universal shape of the echoes, we 
notice that they consist of characteristic pairs of negative and positive extrema of the invariant $\zeta$ appearing on the $z$ axis separated by a time-like spacetime interval, where the negative peak of $\zeta$
of the strongest echo determines $I_K^{\rm max}$ in Fig. \ref{fig:FigIKofA}. Then we can consider the dependence of $\zeta$ on the proper time $\tau$ along the geodesic $x^\alpha(\tau)$ connecting these two nearby spacetime extrema. We multiply both $\tau$ and $\zeta$ by an appropriate power of the same scale factor $\lambda$ which we fix so that the minimal values of the dimensionless function $\zeta_0(\tau_0)\equiv \lambda^2\zeta(x^\alpha(\lambda \tau_0))$ match.

\begin{figure}[t]
    \centering
    \includegraphics[width=\linewidth]{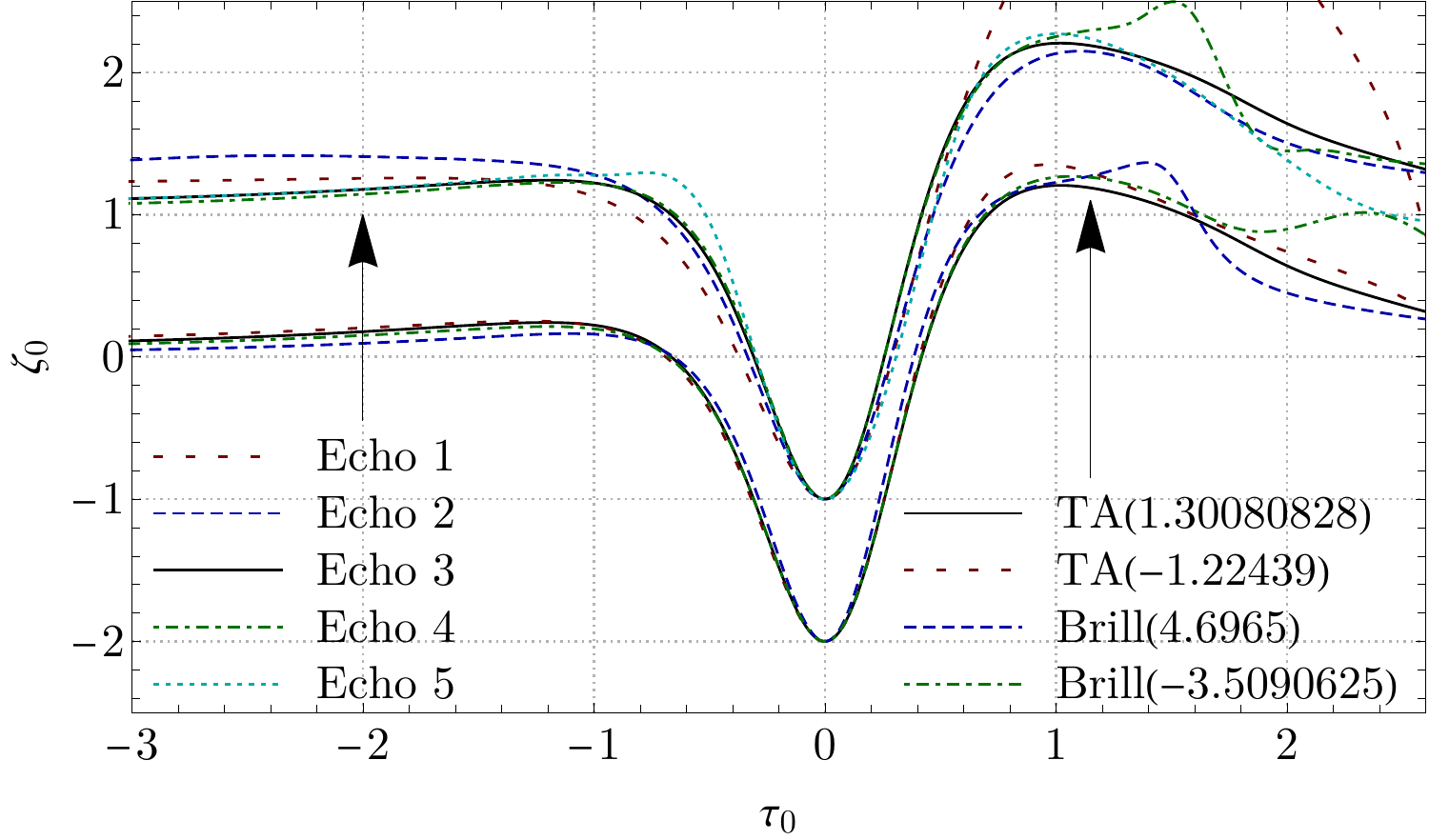} 
    \caption{Profiles of invariant $\zeta$ as a function of proper time $\tau$ along a timelike worldline through ``echo''. To compare the
    profiles, rescaled quantities are used: $\zeta_0=\lambda^2\zeta$ and $\tau_0=\tau/\lambda$, where the same scale $\lambda$ is chosen so that 
    $\min(\zeta_{0})=-2$.
    {\it Top curves} show shifted value $\zeta_0+1$ of five successive ``echoes'' evolved from initial data \eqref{taData} with $A_{\text{TA}^+}=\overline{1.30080828}$. {\it Bottom curves} relate observed profiles of $\zeta_0$ for indicated amplitudes of four different families of initial data.}
    \label{fig:echoes}
\end{figure}

In Fig. \ref{fig:echoes}, we compare such rescaled profiles of the invariant $\zeta$ for different extrema of the same evolution to demonstrate their mutual similarity, and for various initial data families to demonstrate universality. We approximate the geodesic connecting the extrema by the worldline $z=\rm const$, and to draw the curves, we interpolate the grid values by third-order polynomials.
For a generic ID family, the first extreme(s) of $\zeta$ will have a different profile. For Eq. \eqref{BrillInit} with $A>0$, it not only has a different shape, but its amplitude is so high that the next echo does not surpass the already established $I_K^{\rm max}$. As we see in Fig. \ref{fig:echoes}, the profile of this weaker echo already agrees well with that of a  ``universal'' one. Its segment appears in Fig. \ref{fig:FigIKofA} at $\log(\sigma I_K^{1/4})\approx 4$.

A single scalar invariant is not enough to determine the spacetime geometry unambiguously, but because we know that $\zeta$ is the only non-vanishing component of the Riemann tensor at the axis, the echoes also represent approximate scaled copies of the same patch of spacetime. Because near its maximum $\zeta$ changes only slowly in the $z$ direction, it is interesting that a similar but time-symmetric profile of $\zeta$ appears at the axis for the Weber-Wheeler-Bonnor cylindrical GW pulse \cite{Weber1957}.

\vspace{1em}
\section{Conclusions}
Critical collapse of gravitational waves has been studied for a long time with the hope that a clear, universal, discretely self-symmetric structure will appear.
We showed that 
the first echoes in a near-critical collapse
exhibit only a partial similarity to the DSS behavior of a massless scalar field.
While we observed a universal profile of the echo forming patches of strongest spacetime curvature as approximate copies of a universal template, these appear with apparently irregular delays and scales. Thus, we did not observe a universal and regularly self-similar solution in the $A\rightarrow A_*$ limit, and the dimensionless characteristics of the near-critical behavior seem to depend on the ID family.

We think the observed critical behavior, so distinct from that of spherically symmetric fields, requires further attention. 
It is natural to focus on the closest neighborhood of the critical amplitudes, but it is possible that even then the non-universal aspects of the critical GW collapse will remain, because without the spherical symmetry, ID may leave behind a curved spacetime arena in the vicinity of the accumulation event.

As $A\rightarrow A_*$ the numerical simulations become increasingly expensive. It seems important to study the critical behavior of more diverse or more dimensional families of initial data. This may be a computationally cheaper way to understand certain phenomena, e.g., the origins of the apparently irregular echoing structure.

\begin{acknowledgments}
We thank D. Hilditch for interesting discussions, e.g., spotlighting Ref. \cite{PfeifferYork2005}.
This work is supported by the Charles University Project GA UK No. 1176217
and Czech Science Foundation Project No. GACR 21-11268S.
Computational resources were supplied by the project ``e-Infrastruktura CZ'' (No. e-INFRA LM2018140) provided within the program Projects of Large Research, Development and Innovations Infrastructures.
\end{acknowledgments}

\end{thebibliography}

\end{document}